%% file: paper.tex
\documentclass{sigchi}

\pdfoutput=1

\toappear{Publication rights licensed to ACM. ACM acknowledges that this contribution was
authored or co-authored by an employee, contractor or affiliate of the United States
government. As such, the Government retains a nonexclusive, royalty-free right to
publish or reproduce this article, or to allow others to do so, for Government purposes
only. \\
 {\emph{UIST'16}}, October 16--19, 2016, Tokyo, Japan. \\
 \copyright~2016 Copyright held by the owner/author(s). Publication rights licensed to ACM.
ACM ISBN 978-1-4503-4189-9/16/10...\$15.00. \\
DOI: http://dx:doi:org/10:1145/2984511:2984580
}

\usepackage{graphicx}
\usepackage{booktabs}
\usepackage{balance}
\usepackage{enumitem}
\usepackage[super]{nth}
\usepackage{tabularx}

\usepackage{textcomp}
\newcommand{\textapprox}{\raisebox{0.5ex}{\texttildelow}}

\usepackage{url}

\usepackage{color}

\newcommand*\rot{\rotatebox{90}}

\makeatletter
\def\url@leostyle{%
  \@ifundefined{selectfont}{\def\UrlFont{\sf}}{\def\UrlFont{\small\bf\ttfamily}}}
\makeatother
\def\pprw{8.5in}
\def\pprh{11in}

\usepackage[pdftex]{hyperref}
\hypersetup{
pdftitle={SIGCHI Conference Proceedings Format},
pdfauthor={LaTeX},
pdfkeywords={SIGCHI, proceedings, archival format},
bookmarksnumbered,
pdfstartview={FitH},
colorlinks,
citecolor=black,
filecolor=black,
linkcolor=black,
urlcolor=black,
breaklinks=true,
}

\begin{document}

\title{Private Webmail 2.0: Simple and Easy-to-Use Secure Email}

\numberofauthors{1}
\author{
\alignauthor
	Scott Ruoti\raisebox{7pt}{$\dagger$}\titlenote{Sandia National Laboratories is a multi-program laboratory managed and operated by Sandia Corporation, a wholly owned subsidiary of Lockheed Martin Corporation, for the U.S. Department of Energy's National Nuclear Security Administration under contract DE-AC04-94AL85000},
	Jeff Andersen\raisebox{7pt}{$\dagger$},
	Travis Hendershot\raisebox{7pt}{$\dagger$},
	Daniel Zappala\raisebox{7pt}{$\dagger$},
	Kent Seamons\raisebox{7pt}{$\dagger$}\\\affaddr{
		Brigham Young University\raisebox{7pt}{$\dagger$},
		Sandia National Laboratories\raisebox{7pt}{$\ast$}
	}\\
	\email{\{ruoti, andersen, hendershot\} @ isrl.byu.edu, \{zappala, seamons\} @ cs.byu.edu}
}


\maketitle

\input{abstract}

\keywords{Security; Usability; Secure email; Encryption}

\category{H.5.2.}{Information Interfaces and Presentation (e.g. HCI)}{User Interfaces}[user-centered design]  
\category{H.1.2.}{Models and Principles}{User/Machine Systems}[human factors]

\input{intro}

\input{background}

\input{improvements}

\input{method}

\input{results}

\input{discussion}

\input{conclusion}

\section{Acknowledgment}
We would like to thank Mike Jones and Dan Olsen for providing feedback on this paper.

%
\balance

\bibliographystyle{acm-sigchi}
\bibliography{paper}

\end{document}

%% file: abstract.tex
\begin{abstract}
  Private Webmail 2.0 (Pwm 2.0) improves upon the current state of the art by increasing the usability and practical security of secure email for ordinary users.  More users are able to send and receive encrypted emails without mistakenly revealing sensitive information.
In this paper we describe four user interface traits that positively affect the usability and security of Pwm 2.0.
In a user study involving 51 participants we validate that these interface modifications result in high usability, few mistakes, and a strong
understanding of the protection provided to secure email messages.
We also show that the use of manual encryption has no effect on usability or security.

\end{abstract}

%% file: intro.tex
\section{Introduction}
Recent measurements demonstrate that email is largely insecure~\cite{foster2015security,durumeric2015neither,holz2016tls}.
Although adoption of connection-based encryption and authentication is increasing (i.e., STARTTLS, SPF, DKIM, DMARC), these technologies are often configured incorrectly and have weaknesses that render them susceptible to attacks.
However, even if transmission of email were properly secured, email at rest is generally unencrypted.
This means that an honest-but-curious email provider or a 
government that coerces an email provider into granting access
can potentially scan email messages.
End-to-end encryption is a much stronger protection mechanism.
With end-to-end encryption, the sender encrypts email messages before transferring them to the email provider, ensuring that no one but the intended recipients is able to read the messages.

Unfortunately, usable secure email is still an open problem more than 15 years after Whitten and Tygar's seminal paper, {\it Why Johnny Can't Encrypt}~\cite{whitten1999why}.\footnote{We note that S/MIME is widely used in certain organizations (e.g., U.S. government), but this adoption has not spread to the masses.}
Arguably the most secure form of end-to-end encryption is PGP, but PGP has continually proved itself unusable~\cite{whitten1999why,sheng2006why,ruoti2015johnny}.
More recently, research has shown that by relaxing the security requirements of PGP, it is possible to create usable, secure email~\cite{garfinkel2005johnny,ruoti2013confused,atwater2015leading}.
While these newer systems have lower \textit{theoretical} security than PGP, 
they have higher \textit{practical} security---meaining these new tools allow ordinary users to encrypt their email, something they are largely unable to do with PGP-based tools.

Of these usable, secure email systems, Private Webmail (Pwm) is specifically targeted at helping the masses encrypt their existing
webmail accounts, such as those provided by Google, Yahoo, and Microsoft.
A set of user studies of Pwm demonstrated that it had significantly higher usability than similar secure webmail systems (Encipher.it and Voltage Mail~\cite{ruoti2013confused}).
Still, in this evaluation of Pwm about a tenth of participants accidentally sent email in the clear when they meant to encrypt it. Many more users were unsure of the security provided by Pwm.

In this paper, we describe an improved interface design for Pwm that addresses issues raised in the original studies.
First, we added an artificial delay to encryption that enhances users' confidence in the strength of the message encryption while simultaneously instructing users on who can read encrypted messages.
Second, we modified the email composition interface to help users understand which emails are encrypted, so that they avoid mistakenly sending plaintext in the clear.
Third, we added contextual clues that help users understand how to correctly use secure email, showing how the recipient,
subject, and optional greeting fields are used.
Finally, we implemented inline, context-sensitive tutorials, which improved view rates for tutorials from less than 10\% for Pwm to over 90\% in some cases for Pwm 2.0. 
%
Through a user study involving 51 participants, we demonstrate that our new design, referred to as Pwm 2.0, is significantly more usable and secure than Pwm.

Our main contributions are as follows:

\begin{enumerate}

\item
We describe the interface modifications we used to build Pwm 2.0.
Our user study demonstrates that these modifications had a significant impact on the usability and security of Pwm 2.0 as compared to the original Pwm system.
Moreover, we discuss user reactions to these interface modifications and lessons learned from our study.
We believe these principles could also benefit other encryption applications (e.g., instant messaging, social networks, file sharing).

\item
Pwm 2.0 scores an 80.0 on the System Usability Scale (SUS), rating in the ``excellent'' category for usability and receiving an ``A'' grade.
This score is in the \nth{87} percentile for system usability~\cite{sauro2011practical} and is the highest SUS score for secure email to date.
Moreover, over 80\% of users wanted to immediately begin using Pwm 2.0, and over 90\% felt that their friends and family could use Pwm 2.0 with relative ease.

\item
We show that with Pwm 2.0 users rarely send plaintext messages when they meant to encrypt the message.
Out of 306 tasks, only six mistakes were made using Pwm 2.0 (2\%),
a significant improvement to Pwm's rate of about 10\%.
In addition, the majority of users understood the confidentiality, authenticity, and integrity Pwm 2.0 provided their messages (84\%, 63\%, 76\%, respectively).

\item
As part of our study, we split users into two groups
using different versions of Pwm 2.0 supporting either
automatic encryption or manual encryption.
Our results demonstrate that, contrary to the findings of the original Pwm study~\cite{ruoti2013confused}, requiring a user to view their encrypted message before sending it (manual encryption) has no benefit over encrypting and sending the message in one step (automatic encryption).

\end{enumerate}

%% file: background.tex
\section{Background}
In this section, we review related work for secure email and automatic versus manual encryption. 
We then describe the threat model that motivates our work.
Finally, we describe the original Private Webmail system.

\input{related}
\input{threat}
\input{system}

%% file: related.tex
\subsection{Related Work}
Whitten and Tygar~\cite{whitten1999why} conducted the first formal user study of a secure email system (i.e., PGP 5).
They found serious usability issues with key management and users' understanding of the underlying public key cryptography. 
Subsequent studies of newer PGP-based clients have shown that PGP tools are still unusable for the masses~\cite{sheng2006why,ruoti2015johnny}.

Garfinkel et al.~\cite{garfinkel2005johnny} conducted a usability study of secure email based on S/MIME, and observed that hiding encryption details can impact usability. 
For example, they found that because the integration of encryption into Outlook Express ``was a little too transparent,'' users were often unsure whether a given email was encrypted.
Additionally, they found that some users failed to read the instructions associated with various visual indicators, leading to difficulties in understanding the interface.
Still, they found that automating key management increased the overall usability of secure email.

Fahl et al.~\cite{fahl2012helping} explored manual and automatic encryption in the design of a Facebook Chat system.
Here, automatic encryption refers to encrypting and sending a message without explicitly demonstrating that encryption has occurred, whereas manual encryption shows the user ciphertext and requires them to manually send this ciphertext.
Their user study demonstrated that users reported preferring automatic key management, but found no significant difference between automatic and manual encryption. 
Nevertheless, they raised the issue that automatic encryption could impact users' feelings of trust and recommended the issue be addressed in future work.

Ruoti et al.~\cite{ruoti2013confused} conducted a series of user studies with Private Webmail (Pwm), a secure email prototype that tightly integrates with Gmail.
Even though results showed the system to be quite usable, they found that some users made mistakes and were hesitant to trust the system since the automatic encryption was too transparent.
They also conducted a study comparing Pwm to a desktop application (MP) that supported manual encryption.
They found that participants using MP's manual encryption made fewer mistakes and answered more questions correctly on a quiz that tested their understanding of the system.
However, there are significant differences between Pwm and MP, which are confounding factors in Ruoti et al.'s results.
In this paper, we also examine the differences between automatic and manual encryption, but do so using two variants of our Pwm 2.0 system that only differ in regards to the use of automatic and manual encryption.

Atwater et al.~\cite{atwater2015leading} have also examined the question of automatic and manual encryption brought up by Ruoti et al.'s study.
They created two version of a tool based on Mailvelope, and these versions only differed in their usage of manual and automatic encryption.
They found that manual encryption had no statistically significant effect on participants' trust of their secure email system.
Their work differs from ours in that they examined the question of trust whereas we are looking at understanding and mistakes, the two pivotal points from the original Pwm study.
In addition, though Atwater et al. demonstrate usability with a PGP-based client, the system they simulated does not
include several practical key management steps that users would be required to perform, limiting the applicability of its results.

%% file: threat.tex
\subsection{Threat Model}
In our threat model there are four entities:

\begin{enumerate}

\item \textbf{User.}
The user's computer, operating system, and secure email software are considered part of the trusted computing base.

\item \textbf{Webmail provider.}
We consider the webmail provider as an honest-but-curious party.\footnote{An honest-but-curious party will gather any information available to them (e.g., Gmail scans email messages), but will not attempt to break the secure email system (e.g., impersonating the user to the key server) or collude with other honest-but-curious parties).}
The webmail provider has access to the users' encrypted messages, but not to the keys that encrypt those messages.

\item \textbf{IBE Key server.}
Key management in Pwm uses Identity-Based Encryption (IBE)~\cite{shamir1984identity}.
In IBE, a user's public key can be computed using public parameters from the IBE key server and the user's identity (e.g., email address).
To retrieve the private key for a given identity, a user proves ownership of the appropriate identity, and then the key server will generate and send the private key to the user.\footnote{Private keys are not stored on the key server, but rather generated on demand based on a master secret. For added security, this secret can be stored inside of a crypto-card.}

We consider the IBE key server as an honest-but-curious party.
While the key server is responsible for generating a user's private key, it does not have access to the messages encrypted with those keys.

\item \textbf{Adversary.}
The adversary is free to eavesdrop on any communication between users, webmail providers, and key servers.\footnote{In nearly all cases, this communication will be encrypted using TLS, and the adversary only has access to the encrypted packets.}
Additionally, the adversary can attempt to compromise the webmail provider or key server.
The adversary wins if she is able to use these resources to access the plaintext contents of the encrypted email body.

\end{enumerate}

We do not consider attacks directly against the user or trusted computing base (i.e., phishing credentials, installing malicious software).
Similarly, we do not consider an attacker who can compromise fundamental networking primitives (i.e., TLS, DNS).
While these are valid concerns, if the attacker can accomplish these types of attacks,
then they can already do far more damage than they could by breaking the secure email system.
We also note that data needed by the webmail provider to transmit email (e.g., recipient addresses) cannot be encrypted, and may be available to the adversary (e.g., this information is passed over an unencrypted channel).
Our threat model instead focuses on ensuring the data in the encrypted body is safe from an attacker.

To steal the user's sensitive data, the adversary must obtain both the encrypted email and the key material needed to decrypt the email.
The former can be accomplished by either compromising the webmail provider, or intercepting an encrypted email that is not transmitted using TLS.
The latter can be accomplished by compromising the IBE key server.
Just as the adversary must collect the data from both the webmail provider and key server, neither of these parties alone has enough information to unilaterally steal the user's sensitive data.
While as honest-but-curious parties, these two entities will not collude, a government could subpoena both organizations and compromise a user's sensitive data.
If this is a significant concern, it is possible to apply a thresholding scheme to the IBE key server and place the various key servers in different localities that for political reasons will not cooperate.\footnote{Alternatively, the IBE key server could roll over its keys on a regular (e.g., daily) basis, limiting the amount of time an adversary or a government has to steal the user's sensitive data.}

While our threat model is more permissive than the traditional PGP threat model, it is nonetheless useful in a variety of situations.
For example, it satisfies the typical case where a small business or a university has outsourced its email to a third-party service provider such as Google, but does not trust Google with sensitive data.
The business or university does, however, trust a browser extension that it has vetted and a key server that it operates.
Alternatively, our threat model allows every-day users to easily encrypt sensitive email, preventing their webmail providers from storing, scanning, or accidentally leaking sensitive information.
Regardless, the model provides greater assurance than currently available through vanilla SMTP~\cite{durumeric2015neither}.

%% file: system.tex
\subsection{Private Webmail (Pwm)}
Pwm implements secure email through tight integration with Gmail and leverages automatic key management.
When users read or compose secure email, Pwm replaces portions of Gmail's interface with Pwm's own secure interface.
Pwm's interfaces are styled differently than Gmail's, providing a clear demarcation of which information is being protected by Pwm.
Pwm's interfaces are implemented using \textit{security overlays}~\cite{van2009encrypted}.
Security overlays allow Pwm's interfaces to be visually integrated within Gmail's interface, while still protecting the content of Pwm's overlays from Gmail.\footnote{This prevents Gmail from being curious and scanning the contents of encrypted emails.}

All secure email sent using Pwm includes instructions on how to setup Pwm for first-time users.
The setup instructions direct new users to the Pwm website, where they are able to add a bookmarklet to their browser's bookmark storage.\footnote{A bookmarklet is a browser bookmark that contains executable JavaScript. This JavaScript is run on the page that users are viewing when the bookmark is clicked.}
The new user then returns to Gmail and clicks on the Pwm bookmarklet to run Pwm.
The benefits of using a bookmarklet to run Pwm include not requiring installation permission on the machine and also avoiding any user worries related to installing extensions~\cite{ruoti2013confused}.
The drawback is that users are required to click on the Pwm bookmarklet each time they reopen Gmail.

\begin{figure}[t]
\centering
\includegraphics[width=0.8\columnwidth]{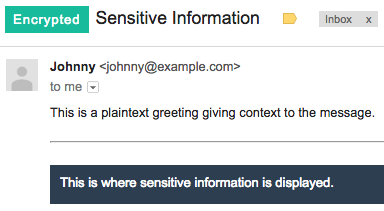}
\caption{The read interface for an encrypted email.}
\label{fig:read}
\end{figure}

\begin{figure*}[t]
\centering
\includegraphics[width=1.0\textwidth]{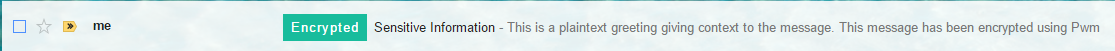}
\caption{An encrypted email in the inbox}
\label{fig:inbox}
\end{figure*}

When Pwm is running, secure email messages are automatically decrypted for users.
The decrypted contents of the message are displayed in place of the instructions and ciphertext that were in the email message's body (Figure~\ref{fig:read}).\footnote{Screenshots in this section are of Pwm 2.0.}
Encrypted email messages in the user's inbox are marked as \textit{Encrypted} (Figure~\ref{fig:inbox}).
Pwm allows users to add encryption on a per-message basis by clicking a lock icon inserted next to the ``Send'' button at the bottom of the compose interface.

In Pwm, users are authenticated to the IBE key server using Simple Authentication for the Web (SAW)~\cite{van2007simple}, a form of email-based authentication \cite{garfinkel2003email}.
Since Pwm runs inside the user's Gmail session, it is able to complete authentication automatically and does not need to prompt the user for input.
It should be noted that because Pwm authenticates to the IBE key server using email, if an adversary were to compromise the user's webmail account they would also be able to authenticate to the key server.
This means that messages encrypted with Pwm are only as secure as the user's webmail account.
Ruoti et al. took this approach in order to maximize usability and because it still provides greater security than currently available to webmail users~\cite{foster2015security,durumeric2015neither,holz2016tls}.
If greater security is needed, the IBE key server could be modified to authenticate users by some other means (e.g., separate password, two-factor authentication).

%% file: improvements.tex
\section{Usability Improvements}
\label{sec:improvements}

We used an iterative design methodology to address problems with Pwm's original interface.
For each problem, we brainstormed several potential solutions and evaluated each solution using cognitive walkthroughs and heuristic evaluations.
The most promising solutions were then implemented in prototypes, and these prototypes were evaluated with pilot usability studies conducted with a convenience sample from friends, family, and students.
Those solutions that were rated as helpful in our pilot usability studies were then implemented in our Pwm 2.0 system.
The source for Pwm 2.0 can be found at \url{https://bitbucket.org/isrlemail/pwm}.

In the remainder of this section, we describe the most relevant solutions that were added to Pwm 2.0.

\subsection{Delayed Encryption}
One concern expressed by a significant portion of participants in the original Pwm study~\cite{ruoti2013confused} was that Pwm encrypted email so quickly that participants were unsure if Pwm had really done anything.
Also, participants indicated the encryption process was so invisible that they were unsure who could read their encrypted email message.

To address both of these problems we added an artificial delay after users click the {\it Send encrypted} or {\it Encrypt} buttons.
For each email recipient, users are shown a message lasting 0.75 seconds that states the email is being encrypted for that user (e.g., ``Encrypting for \texttt{bob@gmail.com}'').
This helps users understand who will be able to read the encrypted email and also lets them feel that something substantial has happened during the encryption process.\footnote{Future work could examine the effect of removing this delay for experienced users who already have a correct understanding of the system's functionality. Additionally, a maximum delay length could be set for large recipient lists.}
Because most email messages have a small number of recipients, this artificial delay does not significantly impact the overall email experience.

\subsection{Compose Interface}
In the original studies, about 10\% of users accidentally sent a sensitive email without first encrypting it.
In each case, the user recognized that they had made a mistake immediately after hitting the send button, but unfortunately their sensitive information had already been transmitted in the clear.
Moreover, many users initially composed their sensitive emails in plaintext, only encrypting them right before sending the message, allowing Gmail to scan their message as it was being typed.

\begin{figure}[t]
\centering
\includegraphics[width=0.9\columnwidth]{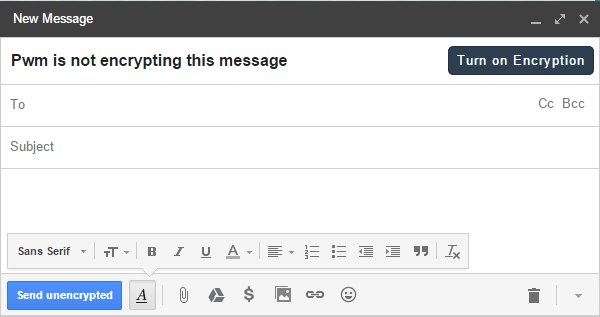}
\caption{Gmail compose interface before enabling encryption}
\label{fig:pre-compose}
\end{figure}

\begin{figure}[t]
\centering
\includegraphics[width=0.9\columnwidth]{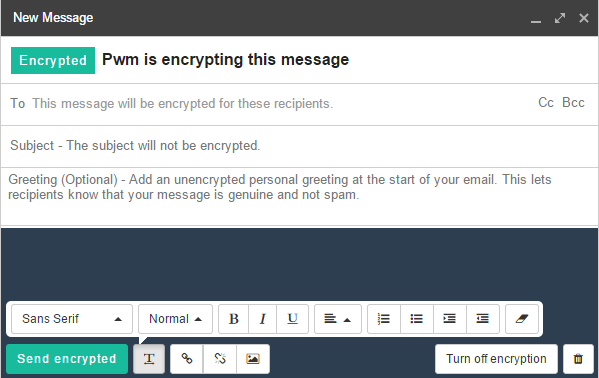}
\caption{Pwm's compose interface}
\label{fig:compose}
\end{figure}

To address both of these issues, we revised Pwm 2.0's compose interface to better conform to the flow of composing email messages in Gmail (i.e., moving from top to bottom).
As part of this process, we added informative text at the top of the compose interface (before the data entry sections) that indicates to users whether their message is being encrypted, then allows them to enable encryption before they begin composing their message (Figure~\ref{fig:pre-compose}).
When encryption is enabled, we modify the informative text to make it clear that the message is now being encrypted (Figure~\ref{fig:compose}).
Furthermore, we modified Gmail's {\it Send} button to read {\it Send unencrypted}, to make it clear to users that by default their messages are sent without encryption.

To help users better understand how Pwm 2.0 protects their emails, we modified the \textit{placeholder} text for the {\tt Recipients} and {\tt Subject} fields, explaining how Pwm 2.0 uses these two fields (Figure~\ref{fig:compose}).
Also, we added functionality allowing users to compose plaintext greetings that are included with the encrypted email (Figure~\ref{fig:read} and Figure~\ref{fig:compose}).\footnote{This feature was first suggested by Ruoti et al.~\cite{ruoti2013confused}, but was not actually implemented.}
The greeting can help engender trust in a new user that receives an encrypted email, giving them confidence to setup and use Pwm 2.0.
It also provides context for encrypted email messages before they are decrypted.

\subsection{Tutorials}
\begin{figure}[t]
\centering
\includegraphics[width=0.6\columnwidth]{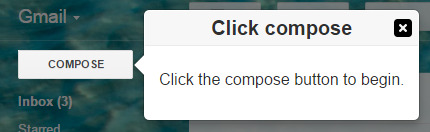}
\caption{Tutorial waiting for action from the user}
\label{fig:tutorial2}
\end{figure}

Several of the problems encountered by participants in the original studies could have been avoided if they had viewed the included video tutorial.
As participants were clearly uninterested in reading the instructions or watching a video on the Pwm website, we replaced these with integrated, context-sensitive tutorials in Pwm 2.0.
These tutorials provide step-by-step instructions on how to use Pwm 2.0, are displayed the first time a user performs a new action, and are shown directly to the side of the interface the user is currently using.
Each step in the tutorials uses simple language and instructs the user about a single feature of Pwm 2.0 (Figure~\ref{fig:tutorial1}).
Some tutorial steps require explicit action from users before they can move on to the next step, helping users internalize correct behavior (Figure~\ref{fig:tutorial2}).
The tutorials are as follows:

\begin{figure}[t]
\centering
\includegraphics[width=0.4\columnwidth]{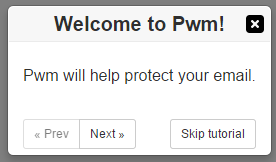}
\caption{Style of the tutorial window}
\label{fig:tutorial1}
\end{figure}

\begin{enumerate}

\item {\bf Introduction.} This tutorial is shown to users the first time they run Pwm 2.0.
It informs users that Pwm 2.0 will help protect their email and tells them how to identify whether Pwm 2.0 is running.

\item{\bf Read.} The first time users read an encrypted email message, they are shown this tutorial.
The tutorial shows users how to identify an encrypted email, explains the plaintext greeting, and identifies which portions of the email message were encrypted.
Additionally, it informs them that email messages encrypted with Pwm 2.0 are provided authenticity and confidentiality (even from Gmail).

\item{\bf Compose.}
The first time users compose an email message while Pwm 2.0 is running, they are given the option to watch a tutorial describing how to compose encrypted email.
This tutorial teaches users the correct order of operations for composing an encrypted email.
It also informs them about who can read their message, the purpose of the optional greeting, and where to type sensitive information.

\end{enumerate}

\subsection{Look and Feel}
We designed a new website for helping users install Pwm 2.0.\footnote{\url{https://pwm.byu.edu}}
This website has a more professional look and feel than the original Pwm website.
This change was made in reaction to user comments that they decide how much they trust a tool based on how professional the tool's website looks.
To make Pwm 2.0 also feel more professional, we standardized its look and feel to use the same colors and styles as the website.

\subsection{Automatic and Manual Encryption}

\begin{figure}[t]
\centering
\includegraphics[width=0.9\columnwidth]{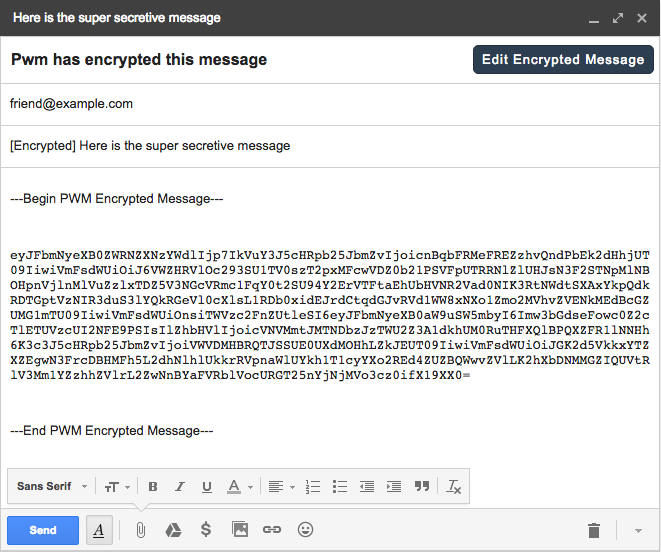}
\caption{Ciphertext shown to users after manual encryption}
\label{fig:pwm-encrypted}
\end{figure}

The original version of Pwm automatically encrypts and sends email as soon as the user clicks {\it Send encrypted} (Figure~\ref{fig:compose}).
We created a manual encryption version that splits this operation into two distinct steps.
In the first step, instead of clicking the {\it Send encrypted} button, the user instead clicks the {\it Encrypt} button.
Upon doing so, the user's email message is encrypted, and both the instructions for decrypting the email message and the encrypted ciphertext are shown to the user inside Gmail's compose interface (Figure~\ref{fig:pwm-encrypted}).
This allows the user to confirm that encryption has actually taken place.
In the second step, the user then clicks Gmail's {\it Send} button to send the encrypted email.

%% file: method.tex

\section{Methodology}
This section gives an overview of our IRB-approved user study that evaluated Pwm 2.0.
This study was also used to run a between-subjects A/B test comparing automatic and manual encryption.
The data from this study is available at \url{https://uist2016.isrl.byu.edu}.

We were careful to design the study to test a generic secure email system, rather than specifically testing Pwm 2.0.
This approach helps us to feel more confident that participants were interacting with Pwm 2.0 in a realistic fashion.
To help with this goal, we included researchers who are not involved in our secure email research in designing the scenarios and tasks.
After creating scenarios and tasks that were acceptable to all parties, we then conducted a pilot study with three participants.
We did not identify any significant flaws during the pilot study.

\subsection{Study Setup}
The study ran for two weeks (February 17, 2015 through March 3, 2015) and included 52 participants that were randomly assigned to test either the automatic or manual encryption version of Pwm 2.0.\footnote{Our power analysis ($\alpha=0.05$, $\beta=0.80$) indicated that we would need 23 participants in each treatment group to reasonably detect differences in SUS scores and task completion time. In contrast, our power analysis revealed that detecting even medium sized (10\%) differences in our proportional measures would require 387 participants in each treatment group (far beyond our group's ability to recruit participants). As such, the proportional measures should only be evaluated as trends, and not as statistically significant.}
Participants took 30 to 60 minutes to complete the study and were compensated \$10 USD for their efforts.

Studies were conducted in a room dedicated for this study.
When participants first entered the room, they were read a brief introduction to the study.
All subsequent instructions were written and provided either via a printed information sheet or via email.
Participants completed all tasks on a virtual machine, ensuring that the computer started in the exact same state for all participants.
We also recorded participants' screens.

Two study coordinators were involved in the study.
One coordinator sat in the same room as the participant, and 
was instructed to avoid prompting participants and to only assist participants if they had not made any progress within five minutes.
The second coordinator sat in another room and corresponded with the participant over email as part of the study tasks.

\subsubsection{Quality Control}
One participant was unable to use Pwm 2.0 with their Gmail account.
This participant had a special Gmail configuration that was not compatible with our tool.
While this participant did complete the study using a temporary email account we provided them, we were worried that their increased interactions with the study coordinator might bias their responses, and so we discarded their results.
For the remainder of this work, we report results based on the 51 remaining participants.\footnote{After analyzing the remaining data, we reviewed the removed participant's results and found that they were in line with the results we analyzed.}

\subsubsection{Participant Demographics}
We recruited Gmail users for our study at a local university.
Participants were evenly split between genders: male (25; 49\%), female (26; 51\%).
Participants skewed young: 18--24 years old (45; 88\%), 25--34 years old (3; 6\%), 35--44 years old (2; 4\%), 55 years or older (1; 2\%).

We distributed posters broadly across campus to avoid biasing our results by any particular major.
All participants were affiliated with the university,\footnote{We did not require this affiliation.} with the overwhelming majority being undergraduate students: undergraduates (44; 86\%), graduates (2; 4\%), faculty and staff (5; 10\%).
Participants had a variety of majors, including both technical and non-technical majors.
No major was represented by more than three participants, with the vast majority having only a single participant.

Most participants indicated they logged into Gmail on the web to check their email frequently:
many times a day (39; 76\%), once a day (3; 6\%), 2--3 times a week (2; 4\%), once a week (2; 4\%), 2--3 times a month (2, 4\%).

\subsection{Scenario and Task Design}
\label{sec:scenariodesign}
During the course of the study, participants were given two scenarios to complete: (1) being hired for a new job, and (2) sending
credit card information to a spouse.
Prior to beginning each scenario, participants were provided with a written description of the scenario and
information that participants should send in place of their own personal information.\footnote{We took this approach to avoid situations where sensitive information might have been leaked to Gmail.}
Participants were asked to treat this sensitive information with the same care as they would treat their own information.

For each scenario we designed realistic tasks that used as many email features as possible.
Participants completed tasks in the order shown below.
Tasks were completed entirely using email, and participants used their own email accounts.
If participants accidentally sent sensitive information without encryption, they were notified of their mistake in an email and asked to resend the information using encryption.

\paragraph{Being hired for a new job}
Participants were told that they had recently returned from an interview at National Citadel, a fictitious company created for this study.
\begin{itemize}[label=,leftmargin=0pt]
	\item \textbf{Task 1.}
	Participants received an email from National Citadel containing instructions on how to be reimbursed for expenses from their recent interview. Participants were told to send their Social Security number and a picture of their receipts. They were informed that this information must be encrypted as per National Citadel's policies. This email also instructed users to set up and use Pwm 2.0 to encrypt their email messages.
	
	This task was designed to test whether participants were able to set up and use Pwm 2.0 using only instructions that might be reasonably expected from a company requesting information to be encrypted.
	
	\item \textbf{Task 2.}
	Participants were asked to simulate several weeks passing after the completion of task by closing their browser and then reopening Gmail. Participants were then sent an encrypted email that contained an offer letter. They were asked to reply with their acceptance and to \textit{CC} their new manager.
	
	This task was designed to test whether participants would remember how to enable Pwm 2.0. It also tested whether they could use Pwm 2.0 to \textit{CC} a new recipient.

	\item \textbf{Task 3.}
	Participants were sent an email instructing them to email information to a background check company. They were instructed to encrypt the information.

	This task was designed to test whether users could enable encryption, either by forwarding the request for information or by composing a new email message.

	\item \textbf{Task 4.}
	Participants were instructed to send bank account information to National Citadel's payroll department. They were {\em not} reminded to encrypt this information.
	
	This task was designed to test whether users would remember to encrypt information if they were not explicitly prompted to do so. Unlike the preceding tasks, if they sent information without encryption, we still considered this task complete.
\end{itemize}

\paragraph{Sending credit card information to a spouse}
Participants were told that their spouse had texted them asking for login information to a credit card website. 

\begin{itemize}[label=,leftmargin=0pt]
	\item \textbf{Task 5.}
	  Participants were instructed to send login information to their ``spouse'' using encrypted email. Participants were told that their spouse had never before used secure email, so they should take whatever steps they felt necessary to ensure their spouse could read the encrypted email.
	
	This task was designed to see how participants would induct a new person into using secure email. Regardless of what instructions were sent, we considered this task complete when the information had been sent.\footnote{Pwm 2.0 includes instructions by default, and since they were sufficient to help the participant start using Pwm 2.0, it was reasonable to assume that the participant believed they would be sufficient to help their spouse start using Pwm 2.0.}
	
	\item \textbf{Task 6.}
	Participants received another email from their spouse asking them for additional credit card information. This request was {\em not} encrypted and did {\em not} instruct the participant to send the additional credit card information encrypted.
	
	This task was designed to test whether users would remember to encrypt information if they were not explicitly prompted to do so. Unlike most of the preceding tasks, if they sent information without encryption, we still considered this task complete.
\end{itemize}

Using the video recording of participants' screens, we measured how long they took completing each task and how often they sent sensitive email without encryption (i.e., made a mistake).

\subsection{Study Questionnaire}
We administered our study using the Qualtrics web-based survey software.
Before beginning the survey, participants were read an introduction by the study coordinator and asked to answer a set of demographic questions.
Participants then completed the six study tasks, following which they were asked to complete a questionnaire regarding their experience.

To measure perceived usability, 
we had participants complete the ten System Usability Scale (SUS) questions \cite{brooke1996sus,from2013sus}.
Answers to these questions are used to derive each version's SUS score, a single numeric score from 0, the least usable, to 100, the most usable, that provides a rough estimate of the version's overall usability.
Recent research has shown that SUS scores are effective for comparing systems across different study populations
and that SUS gives more reliable results than other usability metrics \cite{tullis2004comparison,sauro2012quantifying,ruoti2015authentication}.

After completing the SUS questions, participants were asked several questions regarding whether they would want to use Pwm 2.0 in their day-to-day lives.
We then asked participants questions to examine how well they understood the cryptographic properties of Pwm 2.0.
To test understanding of confidentiality, they were asked to indicate which parties could read a message encrypted with Pwm 2.0.
Similarly, participants were asked whether Pwm 2.0 provided authenticity and integrity for secure email messages composed with Pwm 2.0.
Each question was asked in language that would be approachable to users and did not require a technical background.
Participants were given the option to indicate that they were unsure whether a given property was provided by Pwm 2.0.

\subsection{Limitations}
While our study included students with a diverse set of majors and technical expertise,
it may not be representative of non-student populations.
Likewise, Gmail users may also not be representative of the general population's preferences regarding secure email.
Our study is short-term and is not necessarily representative of how participants would use secure email over a longer period of time.
Further studies could address these limitations.

Our study is a lab study and has limitations common to all studies run in a trusted environment \cite{milgram1978obedience,sotirakopoulos2010did}.
While there are indications that some participants treated the provided sensitive information as they would their own (e.g., refusing to email the provided social security number), there is no guarantee that participants' reactions mimic their real life behaviors.
Additionally, our studies did not test participants' ability to resist attacks.

%% file: results.tex
\section{Results}
In this section we report the quantitative results from our user study.

\begin{table}
\centering

%
%
%
%
%
%

\begin{tabular}{l|c|cc|c|}

	\rule{0pt}{11ex} & \rot{Count} & \rot{Mean} & \rot{\shortstack[c]{Standard\\Deviation}} & \rot{\shortstack[c]{Confidence\\Interval\\($\alpha=0.05$)}} \\
	\midrule
	
	Automatic	& 27 & 79.1 & 9.6 & $\pm$3.69 \\ 
	Manual		& 24 & 81.1 & 9.0 & $\pm$3.60 \\ 
	\midrule	
	
	Overall		& 51 & 80.0 & 9.2 & $\pm$2.52 \\ 
	
	\midrule
	
	Original Pwm Study & 72 & 74.2 & 13.7 & $\pm$3.16 \\
	
  \bottomrule
\end{tabular}

\caption{SUS scores}
\label{tab:sus}
\end{table}


\subsection{Usability}

We evaluated Pwm 2.0 using the System Usability Scale (SUS).
The automatic encryption version of Pwm 2.0 had a SUS score of 79.1 and the manual encryption version had a SUS score of 81.1.
Further breakdown of the SUS scores, along with SUS scores from the original Pwm study ~\cite{ruoti2013confused}, can be found in Table~\ref{tab:sus}.
The difference between manual encryption was not statistically significant (two tailed student t-test, unequal variance---$p = 0.43$).
The difference between Pwm 2.0's overall SUS score (80.0) and Pwm's SUS score (74.2) is statistically significant (two tailed student t-test, unequal variance---$p < 0.01$).

To give greater context for Pwm 2.0's SUS score, we leverage the work of several researchers.
Bangor et al. \cite{bangor2009determining} analyzed 2,324 SUS surveys, and derived a set of acceptability ranges that describe whether a system with a given score is acceptable to users in terms of usability.
Bangor et al. also associated specific SUS scores with adjective descriptions of the system's usability.
Using this data, we generated ranges for these adjective ratings, such that a score is correlated with the adjective it is closest to in terms of standard deviations.
Sauro et al. \cite{sauro2011practical} also analyzed SUS scores from Bangor et al. \cite{bangor2008empirical}, Tullis et al. \cite{tullis2004comparison}, and their own data.
They calculate the percentile values for SUS scores and assign letter grades based on percentile ranges.

Using these contextual clues, Pwm 2.0's SUS score of 80.0 is rated as having ``excellent'' usability and given an ``A'' grade.
It is in the \nth{87} percentile for system usability.

\subsection{Task Completion Times}

\begin{table}
\begin{center}
\resizebox{\columnwidth}{!}{

\begin{tabular}{l|c|cccccc|c|}

	\rule{0pt}{11ex} & \rot{Count} & \rot{Task 1} & \rot{Task 2} & \rot{Task 3} & \rot{Task 4} & \rot{Task 5} & \rot{Task 6} &\rot{Total} \\
	\midrule
	
	Automatic	& 27	& 4:06 	& 4:00 	& 3:01 	& 1:20 	& 4:01 	& 0:54	& 17:22	\\
	
	Manual		& 24 	& 5:10 	& 3:48 	& 3:03	& 1:08 	& 3:44 	& 0:50 	& 17:43 \\ 
	\midrule	
	
	Overall		& 51 	& 4:37 	& 3:54 	& 3:02 	& 1:14 	& 3:52 	& 0:52 	& 17:32 \\

  \bottomrule
\end{tabular}
}
\end{center}

\caption{Average task completion times (min:sec)}
\label{tab:times}
\end{table}

For each task, we calculated the average time between when the user could start a task and when they finished that task.
These are shown in Table~\ref{tab:times}.
For Tasks 2--6, and for the study as a whole, any differences between manual and automatic encryption were negligible.
Task 1 had a statistically significant difference (two tailed student t-test, unequal variance---$p = 0.04$),  but we caution against overemphasizing this result.
First, while the difference is significant, the 95\% confidence interval indicates that the actual effect size might be small ($64\pm60$ seconds).
Second, any differences in task completion times are amortized over the remaining tasks.
Finally, a single statistically significant difference could arguably be the result of alpha inflation~\cite{sauro2012quantifying}.

\subsection{Understanding}

%
%
%
%
%
%
%

We asked three questions to determine whether each participant understood the confidentiality, authenticity, and integrity properties provided by Pwm 2.0.
The responses indicated whether each participant correctly understood the principle, had some misunderstanding, or was unsure.
Overall, our data suggests that over 50\% of users would likely understand the security provided by Pwm 2.0 (Adjusted-Wald binomial confidence interval~\cite{sauro2012quantifying}, $\alpha=0.05$---Confidentiality--$83\%\pm10\%$, Authenticity--$62\%\pm13\%$, Integrity--$75\%\pm12\%$).
In all cases, the differences between manual and automatic encryption were not statistically significant.

The proportion of users who understood how Pwm 2.0 was protecting their email was much higher than the rate reported in the original Pwm study.\footnote{We do not calculate statistical significance for these two proportions as they were derived from different questions. Note that the understanding questions for Pwm 2.0 were stricter than those used in the original Pwm study, emphasizing how much better understanding in Pwm 2.0 is likely to be.}
Still, it is interesting that even with tutorials that explicitly instruct participants on these three cryptographic properties, there are still a small number of participants who were unsure how their messages were protected.
This indicates that more work needs to be done to determine how to effectively teach users about these properties.

\subsection{Avoiding Mistakes}
During the study, we recorded all instances of a participant taking an action that deviated from the study parameters.
Using this data, we identified several instances where a participant's actions leaked sensitive information.
These results are reported by task in Table~\ref{tab:mistakes}.

\begin{table}
\begin{center}

\begin{tabular}{l|c|cccccc|c|}

	\rule{0pt}{11ex} & \rot{Count} & \rot{Task 1} & \rot{Task 2} & \rot{Task 3} & \rot{Task 4} & \rot{Task 5} & \rot{Task 6} &\rot{Total} \\
	\midrule
	
	Automatic	& 27	& 0 	& 0 	& 1 	& 0 	& 0 	& 0		& 1	\\
	
	Manual		& 24 	& 1 	& 0 	& 0 	& 1 	& 0 	& 0 	& 2 \\ 
	\midrule	
	
	Overall		& 51 	& 3 	& 0 	& 2 	& 1 	& 0 	& 0 	& 6 \\

  \bottomrule
\end{tabular}
\end{center}

\caption{Number of participants who sent sensitive information without encryption}
\label{tab:mistakes}
\end{table}

During the study, two participants sent sensitive information without ever running Pwm 2.0, and another refreshed the Gmail page but did not restart Pwm 2.0 before sending sensitive information. In all three cases, the mistakes were caused by the misuse of the tool and occurred before participants had ever seen any differences in the two versions of Pwm 2.0. For this reason, they are reported in the overall number of mistakes, but not under automatic or manual encryption.
The differences between mistakes related to manual and automatic encryption are negligible.

Overall the rate of mistakes was extremely low (2\%, 6 tasks out of 306).
This is lower than the mistake rate reported in the original Pwm study of 10\%, and the difference is statistically significant ($N-1$ two-proportion test for comparing two independent proportions--$p=0.001$).

\subsection{Tutorials}
We noted the number of participants that completed each tutorial.
The video recording for one participant's session was corrupted, and we were unable to determine if they had completed the tutorials. 
The data from the remaining 50 participants indicate that participants completed a significant number of tutorials:
introductory tutorial (46; 92\%), tutorial on reading secure email (46; 92\%), tutorial on composing secure email (27; 54\%).
This is in stark contrast to the original Pwm study where nearly all participants ignored both the setup instructions and a video tutorial included on the Pwm website~\cite{ruoti2013confused}.
This result indicates that participants are willing to watch tutorials, though two criteria seem to be crucial: they appear in-page as participants need them and they contain simple and direct wording.\footnote{We believe fewer participants watched the compose tutorial, which appeared the first time participants clicked on Gmail's ``Compose'' button, because it was not clear they needed to see it at this point and because the tutorial drew attention to the option that they could skip it. Better tutorial design could possibly address this issue.}

\subsection{Acceptability of Pwm 2.0}

Overall, participants responded very positively to the prospect of using Pwm 2.0 in their own lives:
82\% of participants agreed with the statement ``I want to start using Pwm''\footnote{In the study, Pwm 2.0 was referred to as just Pwm.} ($81\%\pm11\%$).\footnote{The confidence intervals reported here were calculated using the Adjusted-Wald binomial confidence interval, $\alpha=0.05$.}
73\% of participants agree that ``I would use Pwm with my friends and family'' ($71\%\pm12\%$).
92\% of participants agreed that ``My friends and family could easily start using Pwm'' ($90\%\pm8\%$).
There were no significant differences between participants who used the automatic or manual versions of Pwm 2.0.

Due to the nature of the study, it is likely that participant responses were overly positive, and that the proportions are somewhat high.
Nevertheless, this data suggests that if introduced to Pwm 2.0, a non-negligible number of participants would want to continue using it, and would feel comfortable using it to send encrypted email to their friends and family.

%% file: discussion.tex
\section{Discussion}
In this section we discuss noteworthy items, including participant experiences, opinions, and preferences regarding secure email
and lessons learned from improving Pwm. We refer to participants here as R1--R52, with the number corresponding to the order in which they participated in our study.

\subsection{Automatic and Manual Encryption}
Our results indicate no significant differences between automatic and manual encryption.
As such, we hypothesize that the effect observed in the original Pwm study was due to confounding factors between the two systems being studied (i.e., automatic encryption--Pwm, manual encryption--MessageProtector).\footnote{This hypothesis is supported by similar results in Atwater et al.'s work~\cite{atwater2015leading}.}
Alternatively, it is possible that manual encryption would have benefited Pwm's original implementation, but that the modifications we made while creating Pwm 2.0 were sufficient to provide those same benefits.

\subsection{Pwm 2.0's High Usability}
Pwm 2.0's SUS score of 80 falls in the \nth{87} percentile for system usability~\cite{sauro2012quantifying}, and is the highest score for secure email systems~\cite{ruoti2013confused,ruoti2016we}---Mailvelope (35, \nth{4} percentile),
Tutanota (52, \textapprox\nth{15}),
Encipher.it (61, \textapprox\nth{30}),
Voltage (62, \textapprox\nth{32}),
Virtru (72, \nth{63}),
original Pwm (74, \nth{70})
.\footnote{Neither Message Protector~\cite{ruoti2013confused} or Atwater et al.'s system~\cite{atwater2015leading} had functioning key management, so their scores are not reported here.}

Additionally, most participants understood how Pwm 2.0 was protecting their messages, only rarely made mistakes,
and were interested in using Pwm 2.0 with their friends and family. These positive opinions were also reflected
in numerous positive qualitative responses.
For example, R26, R39, and R42 expressed, respectively,


\begin{quote}
\textit{
``It was very concise and user friendly and did not require esoteric knowledge to operate. I would definitely feel more comfortable sending sensitive information over email if I were using Pwm versus just sending it via an email provider.''
}
\end{quote}

\begin{quote}
\textit{
``I liked how I could encrypt sensitive information like bank account information, credit card, and other things. It wasn't that hard to use. I didn't have to download anything; all I had to do was just save a bookmark and then click on it. It was really easy to use.''
}
\end{quote}

\begin{quote}
\textit{
``I liked how it made encrypting important information so easy. The tutorial was fast and easy. I like that it is easy and convenient to use with day to day emails. I liked that the background was blue so I knew when it was encrypting.''
}
\end{quote}

\subsection{Delayed Encryption}
While we did not specifically collect data regarding users' opinions toward our delayed encryption mechanism, we note that not a single participant complained about encryption being too fast.
This is especially significant when compared to the original Pwm study, where over 33\% of participants self-reported that encryption was too fast and therefore untrustworthy.
The difference between these two proportions is statistically significant ($N-1$ two-proportion test for comparing two independent proportions--$p<0.001$).
While this may not mean that users' concerns regarding the speed of encryption were eliminated, it is strongly suggestive that they were reduced enough not to be a distraction.

\subsection{Mixed mode email}
The optional plaintext greeting that can accompany email encrypted by Pwm 2.0 is intended to help email senders give confidence to recipients who have never used Pwm 2.0 that the email is authentic and not spam.
Surprisingly, during our study we noted that a small, but significant number of users would include a plaintext greeting in many of their encrypted email messages.
For example, in Task~4 eight participants (8; 16\%) including a greeting stating that they were sending their direct deposit information.
Interestingly, this was actually a feature that seven participants (7; 14\%) listed as one of their favorite features of Pwm 2.0.
R12 and R51 stated, respectively,

\begin{quote}
\textit{
``It wasn't rigid, I could write part of a message and have the other parts encrypted if I wanted. It was very clear what was encrypted and what wasn't [...]''
}
\end{quote}

\begin{quote}
\textit{
``That it lets me chose when to encrypt and when not to. It's also nice to have the option of writing a message before the encrypted part so others know it's not spam.''
}
\end{quote}


\subsection{Look and Feel}
Participants indicated that having a color-scheme that was distinct from Gmail helped them more easily understand what information was encrypted and what information was in the clear.
This intuitive understanding potentially helped participants avoid mistakenly entering sensitive information where it would not be protected.
Additionally, participants mentioned that it helped them feel more confident in the system.
Thus it is clear that when designing tightly integrated systems, it is important to have a distinct look and feel.

\subsection{Instructing New Users}
Even though Pwm 2.0 automatically inserts instructions on how to setup and use Pwm 2.0 in every encrypted message, 
most participants were unaware of this. 
During Task~5, participants would often spend several minutes trying to open an old Pwm 2.0 message to grab the instructions, only to have the message immediately decrypted and the instructions disappear.
It would be helpful to make it more clear that these instructions are always included 
or to add an option that allows users to explicitly add these instructions.

%% file: conclusion.tex
\section{Conclusion}
Our modifications to Pwm's interface have significantly increased its usability and security.
In our user study, Pwm 2.0 is rated with an 80.0 SUS score, the highest reported SUS score for secure email systems.
Over 80\% of participants expressed a desire to begin using Pwm, and over 90\% of participants believed that their friends and family could easily start using Pwm.
Pwm 2.0 also helped participants avoid mistakenly sending their sensitive data in the clear.

This work represents an important step toward providing usable, secure email encryption.
While Pwm 2.0 has lower theoretical security than PGP-based systems, it provides higher practical security for an ordinary user.
In studies of PGP-based systems, most users were unable to encrypt their emails~\cite{whitten1999why,sheng2006why,ruoti2015johnny}, but in Pwm 2.0 all users are able to encrypt their emails.
While there is little evidence that PGP-based systems will ever be sufficiently usable for the masses~\cite{marlinspike2015gpg},
techniques from Pwm 2.0 could be applied to PGP-based systems to raise their relative usability.